# Optimizing Ti Substitution for the Enhanced Densification, Ionic Conductivity, and Microstructure of Garnet-Type $Li_7La_3Zr_2O_{12}$ Solid Electrolytes

Neha[1], A. V. Deshpande[1,*], Muktai Aote[1], Abhishek Pradhan[1].

[1] Department of Physics, Visvesvaraya National Institute of Technology, South Ambazari Road, Nagpur, 440010, Maharashtra, India

***Corresponding Author: Dr. (Mrs.) A. V. Deshpande**

Department of Physics, Visvesvaraya National Institute of Technology, South Ambazari Road, Nagpur, Maharashtra, 440010 (India)

E-mail Address: avdeshpande@phy.vnit.ac.in

## Abstract:

Garnet-type lithium lanthanum zirconium oxide $Li_7La_3Zr_2O_{12}$ (LLZO) is a favorable solid electrolyte for all-solid-state Li-ion batteries due to its wide electrochemical stability, compatible ionic conductivity, and good safety. However, further improvement in ionic conductivity is required for its practical applications. In this work, titanium (Ti) is doped into LLZO to enhance its Li-ion transport properties and structural stability. The series $Li_7La_3Zr_{2-x}Ti_xO_{12}$ has been successfully synthesized using conventional solid-state reaction method. The content of Ti has been varied from 0 to 0.20 atoms per formula unit (a.p.f.u). The conducting cubic phase has been confirmed by the X-ray diffraction technique (XRD). Scanning electron microscopy (SEM) and energy dispersive spectroscopy (EDS) have been used for structural analysis, and elemental distribution. Density measurements have been carried out for all the samples. Electrochemical impedance spectroscopy revealed that the high ionic conductivity of $8.08\times10^{-5}$ $Scm^{-1}$ is offered by the $Li_7La_3Zr_{1.9}Ti_{0.1}O_{12}$ sample, which has the lowest activation energy of 0.37 eV. The DC polarization analysis verified that the main contribution to conductivity in the 0.10 Ti sample comes from ions. A one order of magnitude increase in room temperature ionic conductivity is observed for the 0.10 Ti sample, making it a strong candidate for solid electrolyte applications.

## 1. Introduction:

The increasing deployment of renewable energy storage systems, portable electronics, and electric vehicles has driven the need for efficient and high-performance energy storage systems. Li-ion secondary batteries have drawn a lot of interest among many electrochemical storage techniques due to their benefits, which include light weight, low self-discharge rate, stable electrochemical performance, long cycle life, and high energy density [1]. However, conventional Li-ion batteries use liquid organic electrolytes, which raises safety issues, like electrolyte leakage, flammability, and thermal instability during operation, all of which hinder ion conduction. These safety challenges have spurred researchers to develop safer and more reliable battery systems[2]. All-solid-state lithium-ion batteries (ASSLIBs) have attracted significant interest as a possible replacement for traditional lithium-ion batteries. In these batteries, solid electrolytes replace liquid electrolytes and facilitate the conduction of Li- ions between electrodes[3]. Solid electrolytes provide numerous benefits over liquid electrolytes, including enhanced safety, good chemical compatibility with lithium metal anode, improved thermal stability, a broader electrochemical stability window, and are considered environment friendly and cost-effective materials[4–6].

Over the years, various solid electrolyte materials have been explored, including LISICON, NASICON, $Li_3N$ types, LIPON, and beta-alumina, as well as sulfide and thin-film solid electrolytes. However, all of these solid electrolytes are chemically unstable with a lithium metal anode and have a narrow electrochemical potential window (<5V)[7]. Additionally, synthesizing these materials in pure and stable phases is often challenging, which reduces their ionic conductivity and overall electrochemical performance[8,9]. There is a need of a solid electrolyte that can address these issues. Among all the studied solid electrolytes, the inorganic oxide-based electrolyte $Li_7La_3Zr_2O_{12}$, commonly known as LLZO, has attracted significant attention because it possesses the essential qualities required for ASSLIBs[10,11].

Generally, the garnet-structured LLZO exhibit two phases: the tetragonal and the cubic phase. The structural framework of both the phases is same, but the main difference lies in the distribution of Li-ions within the lattice. The tetragonal phase has ordered Li-ion distributions, which provide thermodynamic stability to the structure, limiting its conductivity to $10^{-6}$ to $10^{-7}$ $Scm^{-1}$. Whereas the cubic phase has disordered Li-ion distributions, which enhance its ionic conductivity to $10^{-3}$ to $10^{-4}Scm^{-1}$. However, due to the disordered Li-ion distribution, at room temperature, the cubic phase is not stable compared to the stable tetragonal phase. Therefore, to utilize LLZO in ASSLIBs, it is necessary to ensure the stability of the conducting cubic phase of garnet LLZO at room temperature[12,13].

Two strategies have been developed for stabilization of the desired cubic phase at room temperature. The first one involves introducing dopants in the LLZO lattice structure, and the second approach involves the use of a suitable external sintering additive to lower the temperature needed for achieving the cubic phase[8,14–16].

In substitutional doping, $Li^+$, $La^{3+}$, and $Zr^{4+}$ sites in LLZO can be doped with subvalent, isovalent, and supervalent dopants. In order to synthesize the stable cubic phase of LLZO and enhance ionic conductivity at ambient temperature, numerous studies have been conducted to determine the suitable dopants for the appropriate site. According to studies, substituting the Li site with Al and Ga can give the best outcomes [7]. Other than this, Ge, Be, and Zn are also explored as dopants at the Li site[17–20].

The substitution at the La site in LLZO is not as common as Li site doping. However, studies have shown that different doping cations, such as divalent, trivalent and tetravalent, can replace the La ions in the LLZO lattice. At the site of La, dopants like Ca, Ba, Sr and Y are explored[21,22,31,23–30]. Although Zr is a common element in Li garnets, doping at the site of Zr continues to attract considerable research attention. Numerous studies have been conducted by doping the Zr site with various cation dopants, like Ta[21,32], Mo[33], Sb[34], Ce[35], and Ge[36]. In addition to single site doping, a lot of research is being done on co-doping in LLZO, which involves doping at any of the two sites at once to maintain the stability in the structural framework of LLZO. Furthermore, in case of tri-doping, doping is done simultaneously with multiple dopants at the probable sites, i.e. Li, La, and Zr[12,21,37–39].

In many earlier studies, the conducting cubic phase was only achieved after sintering at high temperatures for a long duration, which significantly affected the Li content. Another important factor affecting electrolyte performance is the microstructure of the material. Microstructural properties such as grain size, grain boundaries, and porosity can significantly influence the ionic conductivity of the electrolyte. A dense microstructure with well-connected grains facilitates Li-ion transport, while poorly connected grains and pores serve as barriers.

A few external additives have been employed as sintering aids to reduce the required sintering time and temperature. According to studies, Ti is a useful dopant in LLZO that can be used as a sintering aid. Furthermore, it has been observed that Ti promotes densification and is helpful in stabilizing the cubic phase. Even though there are studies on dual doping of Ti with elements like Al, In, and Ga, but Ti as a single dopant has not been explored yet[40–42]. Therefore, the present study focuses on the synthesis and investigation of Ti as a dopant in LLZO at the Zr site. The Ti content has been varied, over the range of 0 to 0.20 a.p.f.u., to observe its effect on LLZO.

# 2. Experimental section:

## 2.1. Sample preparation

The samples of the series $Li_7La_3Zr_{2-x}Ti_xO_{12}$ (x=0 - 0.20) were prepared using a conventional solid-state reaction method in ambient atmosphere. High-purity $Li_2CO_3$ (Merck, > 99.9%), $La_2O_3$ (Sigma Aldrich, > 99.99%), $ZrO_2$ (Sigma Aldrich, > 99.9%), and $TiO_2$ (Merck, > 99.9%) were used as precursors. An additional 10% of $Li_2CO_3$ is incorporated to prevent lithium loss during high-temperature processing. All precursor powders were weighed accurately in their stoichiometric proportions and initially dry-mixed in an agate mortar-pestle, then wet-mixed with acetone to ensure homogeneity. The resulting mixture was transferred into an alumina crucible for calcination in a muffle furnace at a temperature of 900°C for 8 hours. Following calcination, the calcined powder was allowed to cool naturally to room temperature and ground into a finely homogenized powder. The calcined powder was pressed uniaxially under 4 tons of pressure using a hydraulic press to make pellets of about 10 mm in diameter and 1.5 mm in thickness. The formed pellets were then transferred into a closed alumina crucible containing the mother powder and sintered for 8 hours at 1100°C.

## 2.2. Characterizations:

The crystal structure and the phase purity of pellets sintered at 1100°C were analysed by XRD employing a RIGAKU diffractometer with Cu-kα radiation (Wavelength: 1.54 Å). The diffraction data were recorded within the range of 10°-70° with a step size of 0.02° and a scanning rate of 2°/min. The Archimedes' principle was used to determine the bulk density of the samples using a K-15 Classic (K-Roy) instrument with toluene as the immersion medium. For microstructural and compositional investigation, a scanning electron microscope (JSM-7600 F/JEOL) with EDX was used. The AC conductivity measurements were determined using a NOVOCONTROL impedance analyzer with a frequency range from 20 Hz to 20 MHz and a temperature ranging from ambient temperature to 150°C. Additionally, the DC polarization approach was employed with a KEITHLEY 6512 programmable electrometer to confirm the predominance of ion conduction within the prepared samples. After sintering the pellets were coated with silver paste to ensure ohmic contact with ion-blocking silver electrodes for AC and DC conductivity measurements. The transport number was calculated from the obtained conductivity data.

## 3. Results and discussion

## 3.1. X-Ray Diffraction Analysis

The X-ray diffraction patterns of samples of the series $Li_7La_3Zr_{2-x}Ti_xO_{12}$ (x = 0 - 0.20) are presented in Fig.1(a). All the prepared samples possessed the cubic garnet phase with space group Ia-3d, which is confirmed from the JCPDS file no.45.0109. The diffraction peaks are indexed with their respective (hkl) planes. Alongside the main cubic phase, the presence of $Li_2ZrO_3$ phase was observed around 21° (JCPDS 041-0324). $Li_2ZrO_3$ has a monoclinic crystal structure with space group C2/c1. It is thermodynamically stable against Li, with good structural stability and a high diffusion coefficient. Moreover, the successful incorporation of Ti into the LLZO structure can be validated by the shifting of diffraction peaks, as shown in Fig.1(b). It is clearly observed that the shifting has occurred towards the higher 2θ values. This shift can be attributed to the substitution of $Zr^{4+}$(0.78Å) by the smaller $Ti^{4+}$(0.60Å) ions [41,43] , causing lattice contraction. However, for x = 0.20 a.p.f.u., the peak is clearly moved towards a lower two theta value. This indicates that, beyond the solubility limit, excess Ti could not be incorporated into the LLZO lattice. Moreover, it can also be confirmed from the obtained lattice parameters for the synthesized samples as shown in Fig.2. Here, with the increase in Ti content, the lattice parameters first decreased up to 0.10 a.p.f.u and then increased with the further addition of Ti.

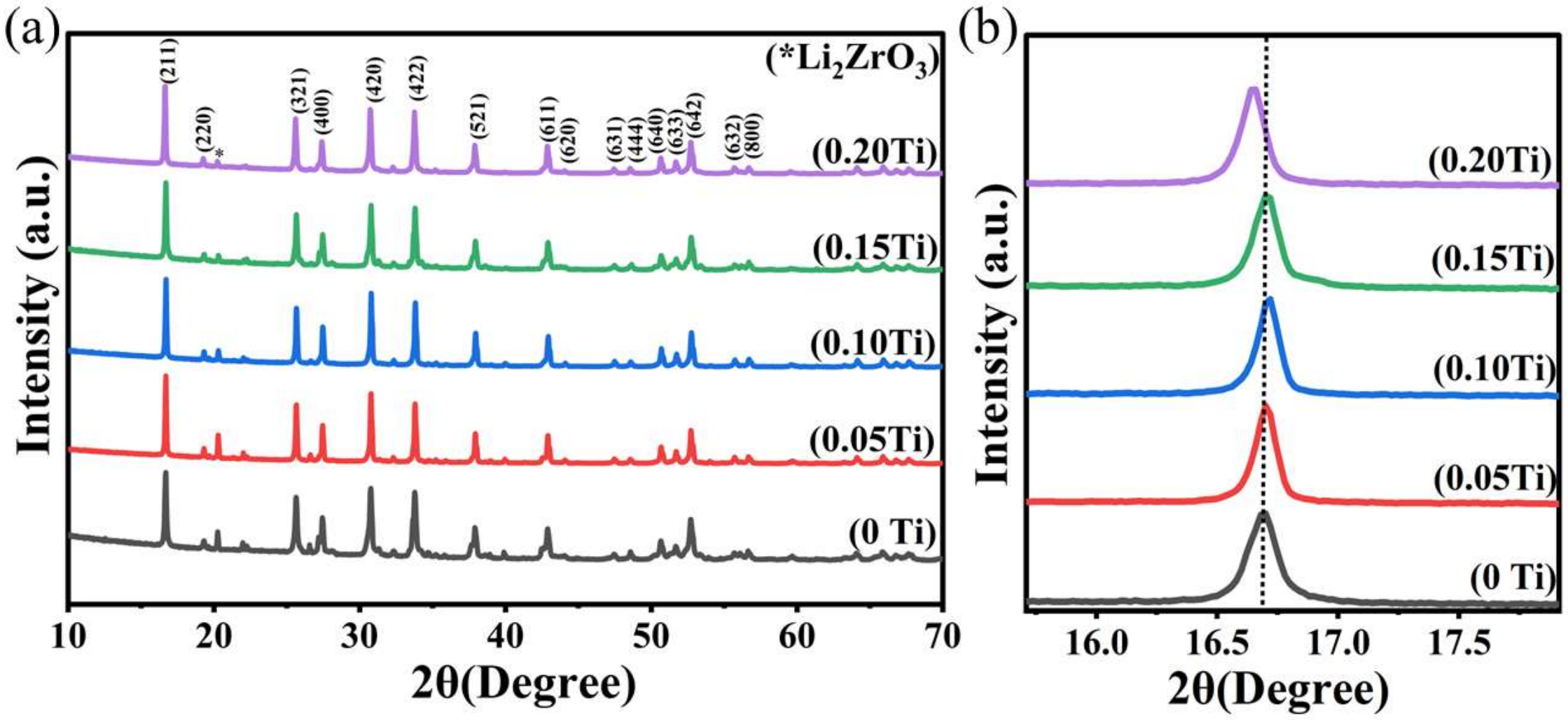

**Fig.1:(a) Powder X-ray diffraction profiles of $Li_7La_3Zr_{2-X}Ti_XO_{12}$ (x = 0 to 0.20) compositions, (b) enlarged (211) peak illustrating the shift in peak position with Ti incorporation.**

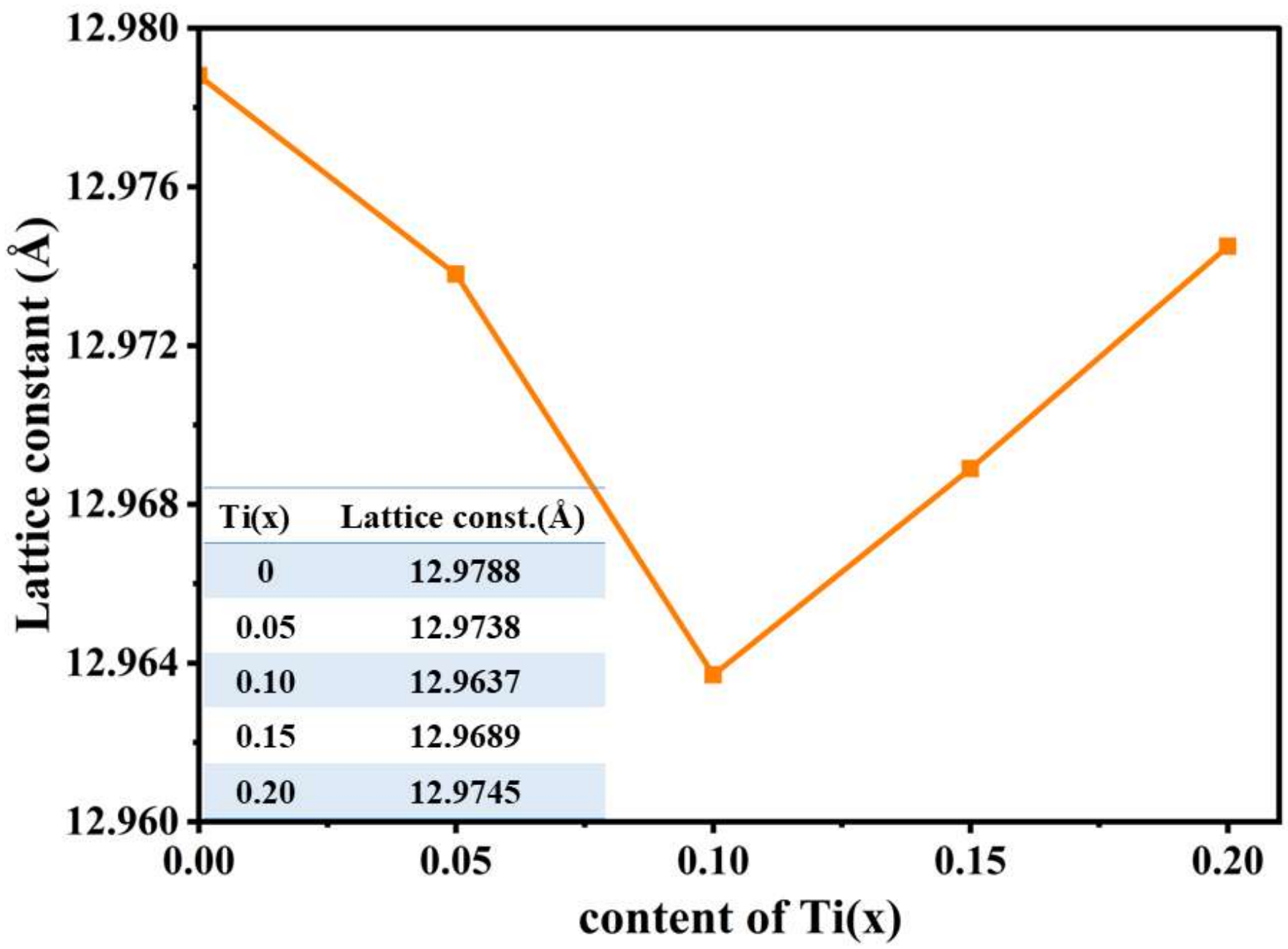

**Fig.2: Lattice constant of the series $Li_7La_3Zr_{2-X}Ti_XO_{12}$ with varying Ti content.**

## 3.2. Density Analysis:

The densities of the series $Li_7La_3Zr_{2-x}Ti_xO_{12}$ (x = 0 - 0.20) have been determined using Archimedes' principle, and toluene is used as an immersion medium. Table 1 represents the bulk density and relative density values obtained for all the prepared samples. It is observed that the density of samples increased with increasing Ti content up to 0.10 a.p.f.u. and then decreased. The sample with x = 0.10 attains the maximum bulk density of 4.69 g/cm³ with the highest relative density of 91.8%. This enhancement in density value can be associated with the suitable amount of Ti incorporation in the LLZO lattice, which helps in forming the compact microstructure[42]. Whereas, with a further increase in Ti content beyond 0.10 a.p.f.u., the density of the samples has decreased, which can be attributed to the development of pores within the structural framework of the samples. This can also be seen in the surface morphology images. However, even though the density is decreased after 0.10 Ti, all the other samples also possessed a relative density higher than 84%, suggesting the sintering behaviour of Ti.

**Table 1. Density and relative density values of the series $Li_7La_3Zr_{2-X}Ti_XO_{12}$(x =0 to 0.20).**

| Ti content | Density ($g.cm^{-3}$) | Relative Density (%) |
|---|---|---|
| 0 | 4.32 | 84.7 |
| 0.05 | 4.56 | 89.2 |
| **0.10** | **4.69** | **91.8** |
| 0.15 | 4.51 | 88.3 |
| 0.20 | 4.42 | 86.5 |

## 3.3. Surface morphology and EDS studies:

Fig.3 (a – e) presents the surface micrographs of the $Li_7La_3Zr_{2-x}Ti_xO_{12}$ series (x = 0 – 0.20). Fig.3(a) shows the microstructure of the undoped sample (x = 0), where a significant number of pores are visible, which tends to decrease the overall density. The addition of 0.05 a.p.f.u. of Ti promotes grain enlargement, which can be observed from Fig.3(b). With the Ti content further increased to 0.10 a.p.f.u., the grains appear well-developed and interconnected with neighbouring grains. No pores can be seen within the sample of this composition, resulting in a dense and compact microstructure, as illustrated in Fig.3(c). This observation is consistent with the highest bulk and relative density values as presented in table 1. Density values indicate that 0.10 Ti promotes effective densification and enhances the microstructural compactness, which facilitates Li-ion migration. When the Ti content exceeds 0.10 a.p.f.u., further grain growth is observed; however, the appearance of voids within the structure also becomes evident, as seen in Fig.3(d,e). This explains the corresponding variation in density values. Therefore, the dense microstructure obtained at 0.10 Ti may be due to the optimum incorporation of Ti into the lattice. The optimal Ti concentration promoted the grain growth in the 0.10Ti sample, which significantly affected the grain boundary migration and ion transport within the LLZO structure[40]. This can also be observed from the conductivity analysis. Furthermore, the average particle size distribution of the 0.10 Ti sample is shown in Fig.4, calculated using ImageJ software, with an average particle size of 8.32μm. The elemental mapping of the 0.10Ti sample, presented in Fig.5, revealed a uniform distribution of La, Zr, Ti and O throughout the surface without any evidence of chemical segregation.

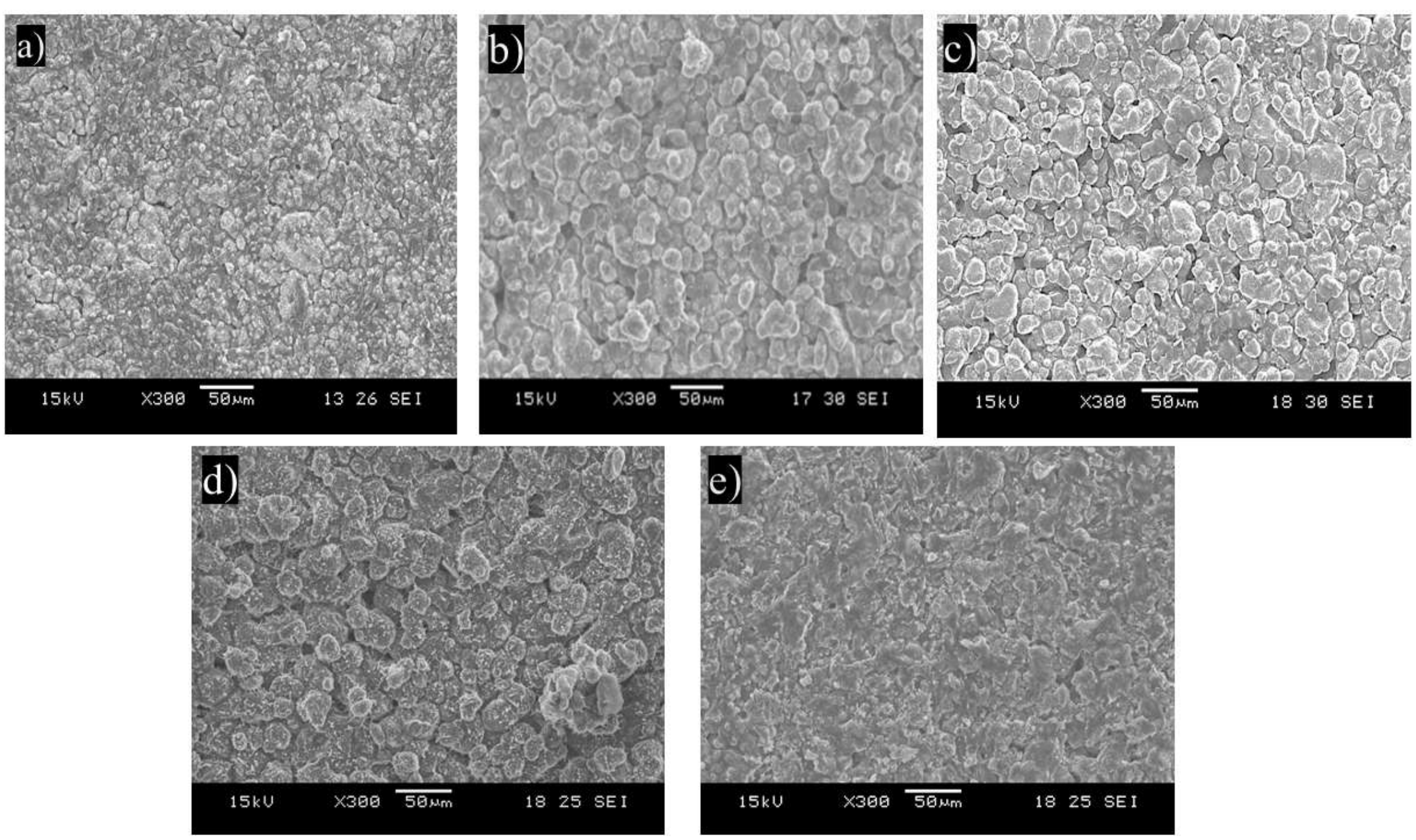

**Fig.3: SEM images of $Li_7La_3Zr_{2-x}Ti_xO_{12}$ with x = a) 0, b) 0.05, c) 0.10, d) 0.15, e) 0.20.**

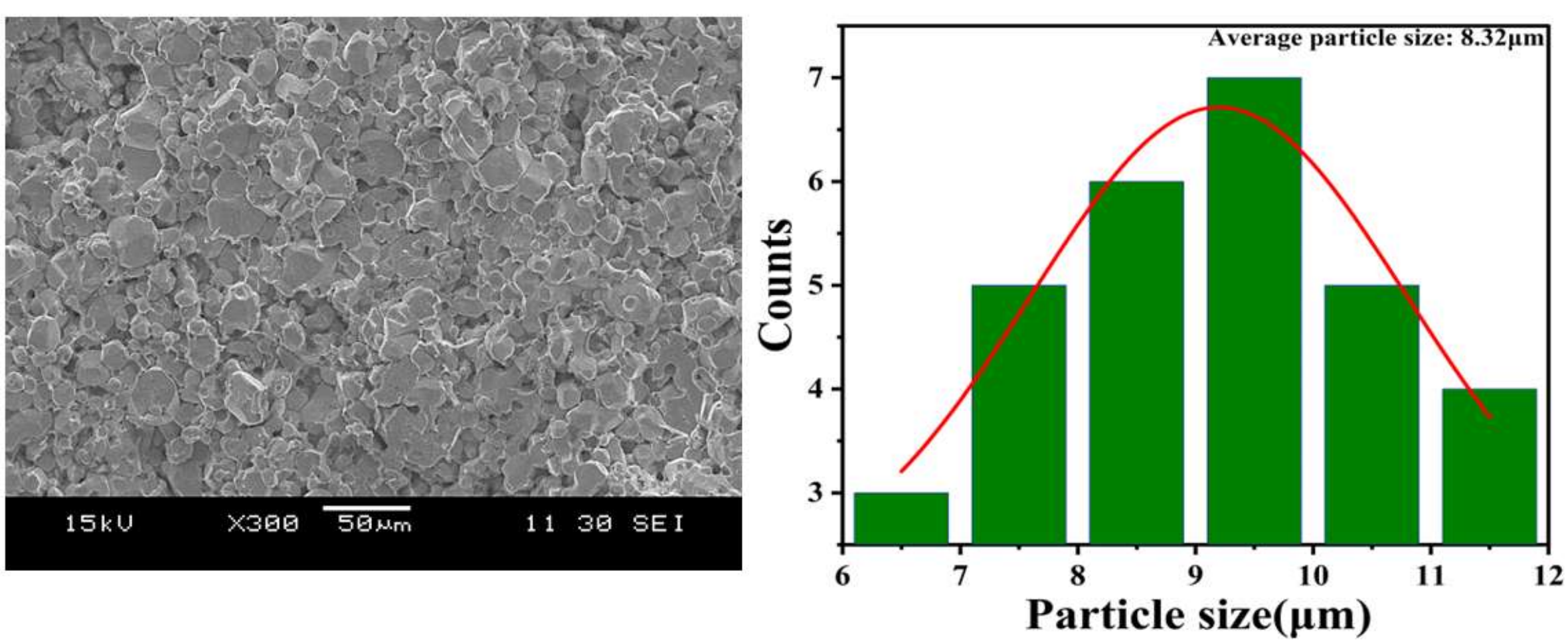

**Fig. 4: Histogram of average particle size distribution of 0.10 Ti sample.**

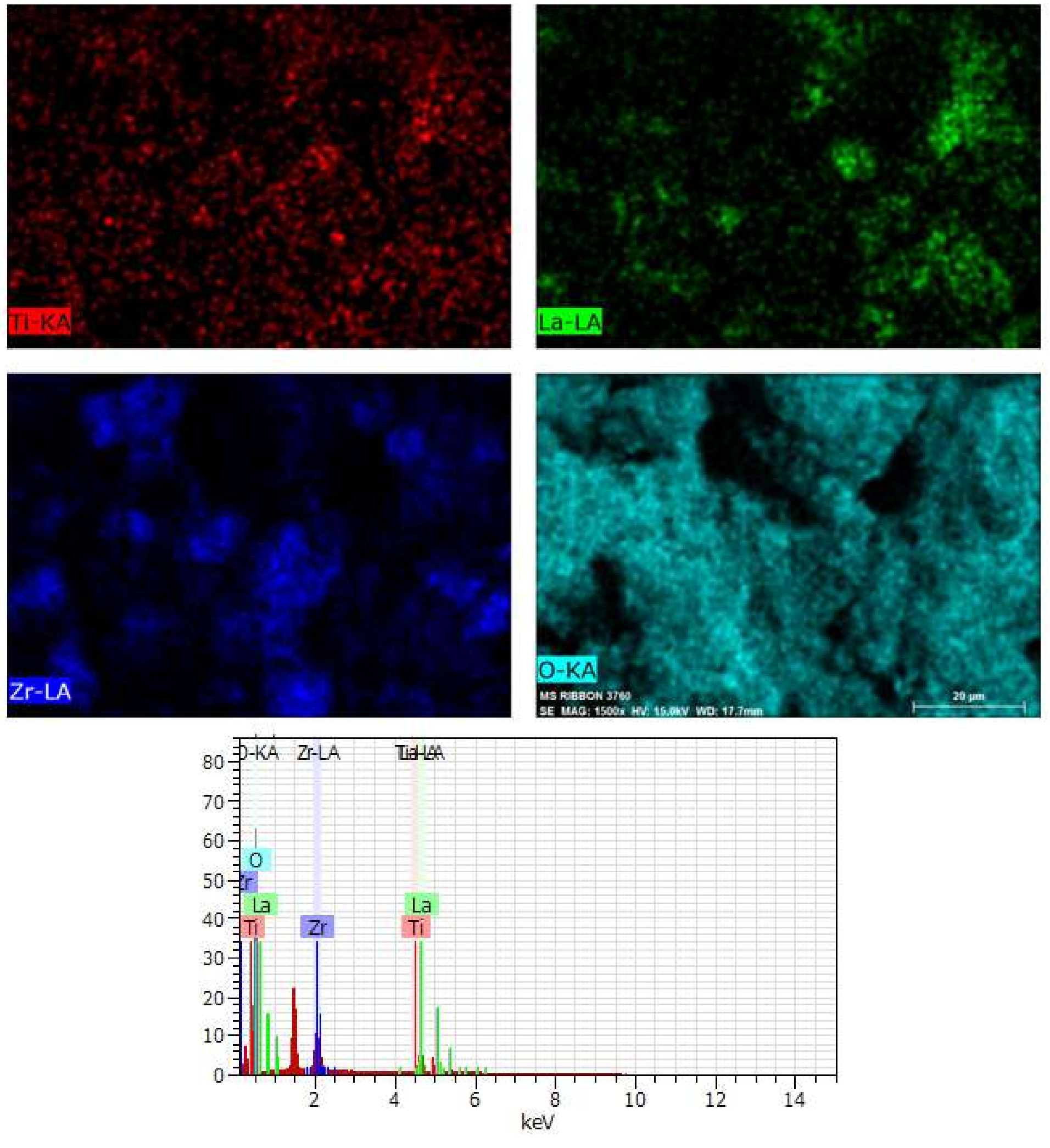

**Fig.5: Elemental mapping and EDS analysis of 0.10 Ti sample.**

## 3.4. AC conductivity studies

### 3.4.1. Nyquist plots

Fig.6(a) illustrates the complex impedance plots of the series $Li_7La_3Zr_{2-x}Ti_xO_{12}$ (x = 0 – 0.20 a.p.f.u.) measured at room temperature. The frequency range in which the impedance data have been collected ranged from 20 Hz to 20 MHz. In general, the Nyquist plot typically consists of a semicircle in the high-frequency region and a tail in the lower-frequency region. The high-frequency semicircle is associated with the combined contribution of bulk and grain-boundary resistances, while the low-frequency tail originates from the polarization effect at the ion-blocking silver electrodes.

The occurrence of a semicircle and a tail indicates that the conduction is mainly ionic. Impedance values can be determined from the intercept of the semicircle on the real z-axis. The total ionic conductivity has been calculated using the relation $\sigma_{total} = t/RA$, where $\sigma_{total}$ is the ionic conductivity, t is the thickness of the sample, R is the resistance offered by the sample, and A is the area of the sample[8,20,35,44].

From fig.6(a), it can be observed that the sample without Ti (0 Ti) exhibits the largest semicircle and, consequently, the highest resistance. However, with increasing Ti content, the semicircles moved towards minimum resistance values. Among all the samples, 0.10 Ti has offered the lowest resistance, offering the maximum ionic conductivity of $8.08 \times 10^{-5}$ $Scm^{-1}$ at room temperature. The fitted impedance spectra of 0.10 Ti is shown in Fig.6(b). The equivalent circuit consists of grain resistance ($R_1$), grain-boundary resistance ($R_2$), grain capacitance ($C_1$), a constant phase element ($CPE_1$), which represents non-ideal grain boundary capacitance, and a Warburg diffusion element ($W_1$) associated with ion diffusion. The highest conductivity is obtained for the 0.10 Ti sample, which can be attributed to the highest relative density and compact microstructure of interconnected grains, which enhances the Li-ion mobility within the structure. However, further increases in Ti content exceeding 0.10 a.p.f.u. lead to a slight increase in resistance and a corresponding decrease in conductivity. This behavior is consistent with the SEM observations as observed in Fig. 3 (d and e), where the appearance of intergranular voids and pores at higher Ti contents is expected to impede ion transport. Thus, the obtained results validated that the optimum Ti content of 0.10 a.p.f.u is beneficial for the improvement in Li ion conductivity.

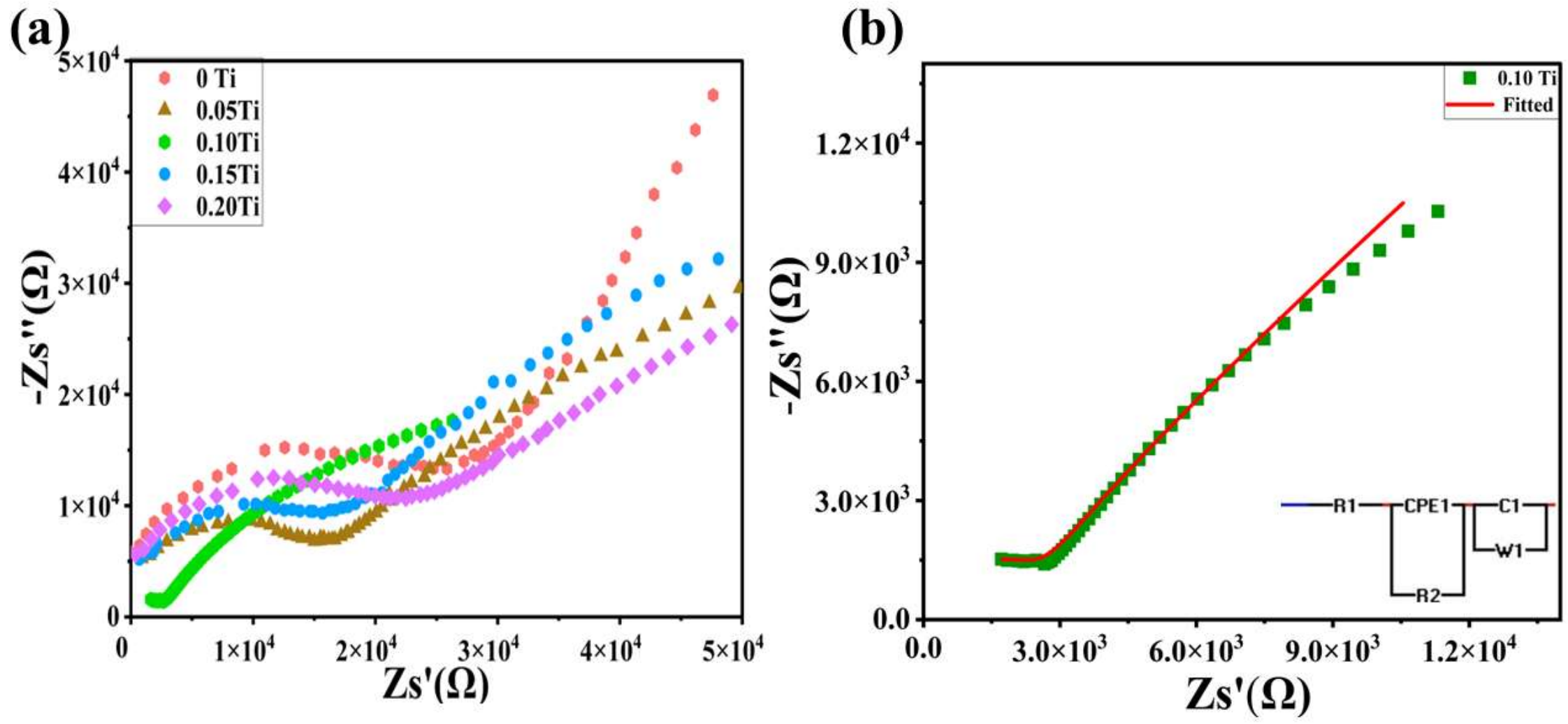

**Fig.6: a) Room temperature impedance plots for the series $Li_7La_3Zr_{2-x}Ti_xO_{12}$ with variation in Ti(x) content from 0 to 0.20 a.p.f.u.**

**b) Fitted impedance spectra and corresponding equivalent circuit of the 0.10 Ti sample.**

## 3.4.2. Arrhenius plots:

Fig.7(a) shows the Arrhenius plots of the series $Li_7La_3Zr_{2-x}Ti_xO_{12}$ with Ti varying from 0 - 0.20 a.p.f.u.. The data was collected at numerous temperatures ranging from ambient temperature to 150°C. Activation energy has been calculated using the Arrhenius equation $\sigma(T) = \sigma_o \exp(-E_a/k_BT)$, where $E_a$ is the activation energy, σ is the conductivity, $\sigma_0$ is the pre-exponential factor, $K_B$ is the Boltzmann constant, and T is the temperature in Kelvin. From Fig. 7(a), it is clearly seen that the conductivity values rise as the temperature rises, which suggests the Arrhenius behaviour in the samples. Among all the samples, the minimum activation energy of 0.37 eV is recorded for the 0.10 Ti sample, which also exhibits the highest ionic conductivity of 8.08 x $10^{-5}$ $Scm^{-1}$. This can be attributed to the reduced lattice constant as mentioned in the XRD results, which may shorten the $Li^+$ ion migration path, thereby facilitating faster ion mobility. The ionic conductivity values along with the corresponding activation energy for the prepared series are summarised in Table 2. In addition, Fig. 7(b) represents the variation in activation energy and room temperature ionic conductivity as a function of Ti concentration in the $Li_7La_3Zr_{2-x}Ti_xO_{12}$ series. Furthermore, the increase in activation energy beyond 0.10 Ti can be due to the appearance of pores and voids within the structure, which eventually hinder the Li-ion mobility.

**Table 2: Ionic conductivity values at room temperature and values of activation energy for the series $Li_7La_3Zr_{2-x}Ti_xO_{12}$.**

| Ti content | Ionic conductivity ($Scm^{-1}$) | Activation energy (eV) |
|---|---|---|
| 0 | $7.51 \times 10^{-6}$ | 0.51 |
| 0.05 | $1.39 \times 10^{-5}$ | 0.41 |
| **0.10** | **$8.08 \times 10^{-5}$** | **0.37** |
| 0.15 | $1.24 \times 10^{-5}$ | 0.42 |
| 0.20 | $9.03 \times 10^{-6}$ | 0.48 |

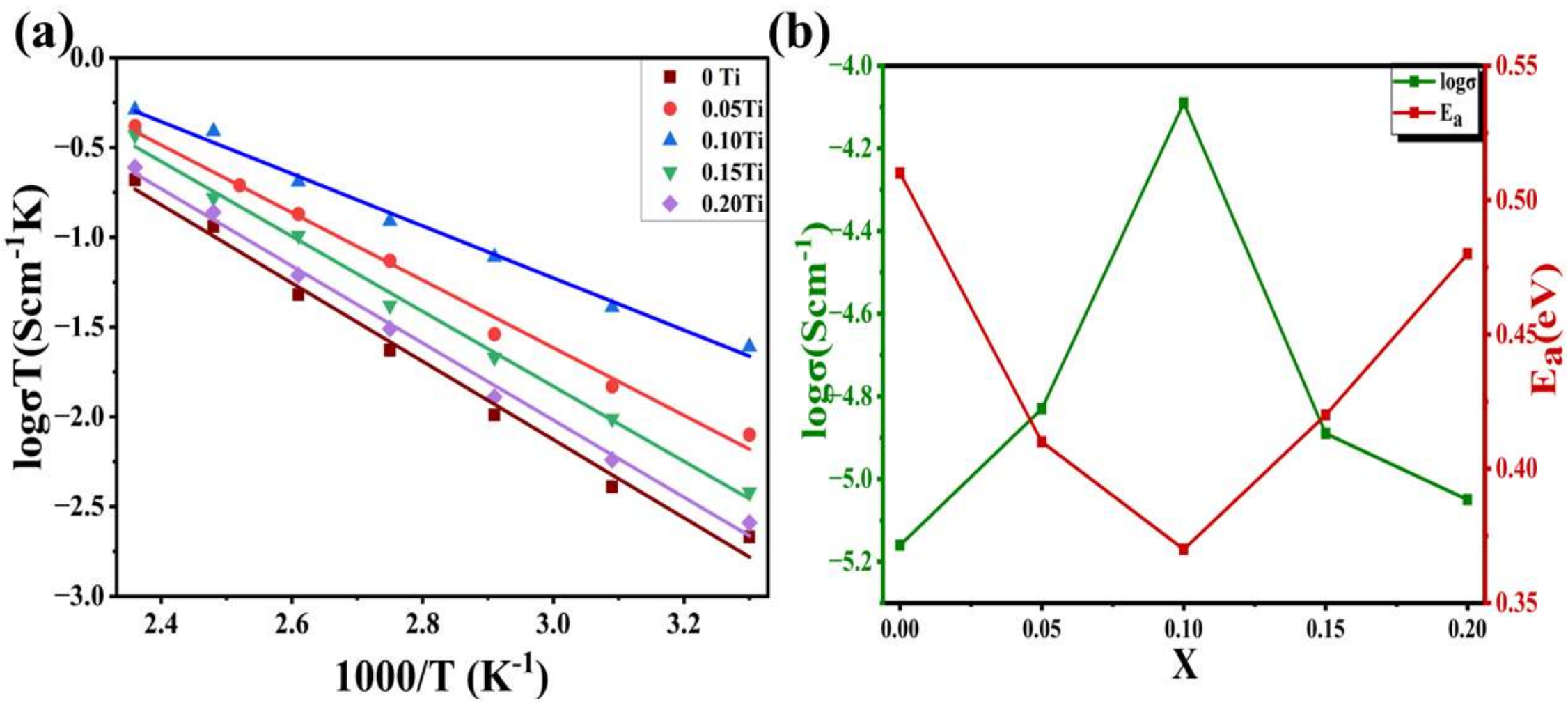

**Fig.7: a) Temperature-dependent ionic conductivity (Arrhenius plots) and b) effect of Ti substitution on the values of room-temperature ionic conductivity and activation energy.**

## 3.5. DC conductivity measurement:

In order to evaluate the role of electronic conductivity in the total conductivity, DC measurement analysis has been performed for all the prepared samples. Fig.8 represents the DC conductivity plot of the 0.10 Ti sample. In this, a similar assembly is used as in the AC conductivity measurement. A constant DC voltage of 1 V has been applied across the pellet, and the resulting current is recorded at equal intervals of time. It is evident from the graph that, after some time, there is a sharp decrease in the current, and it remains almost constant, reaching the steady state value. This stable current observed at longer times is attributed to electronic conduction, because the ionic contribution becomes negligible under steady DC conditions[21,34]. From this, the transport number is calculated using the formula $t_i = (\sigma_{total} - \sigma_e) / \sigma_{total}$. The calculated ionic transport number for 0.10 Ti is obtained to be greater than 0.999, clearly indicating the predominance of ionic conduction within the sample, while electronic conduction is almost negligible.

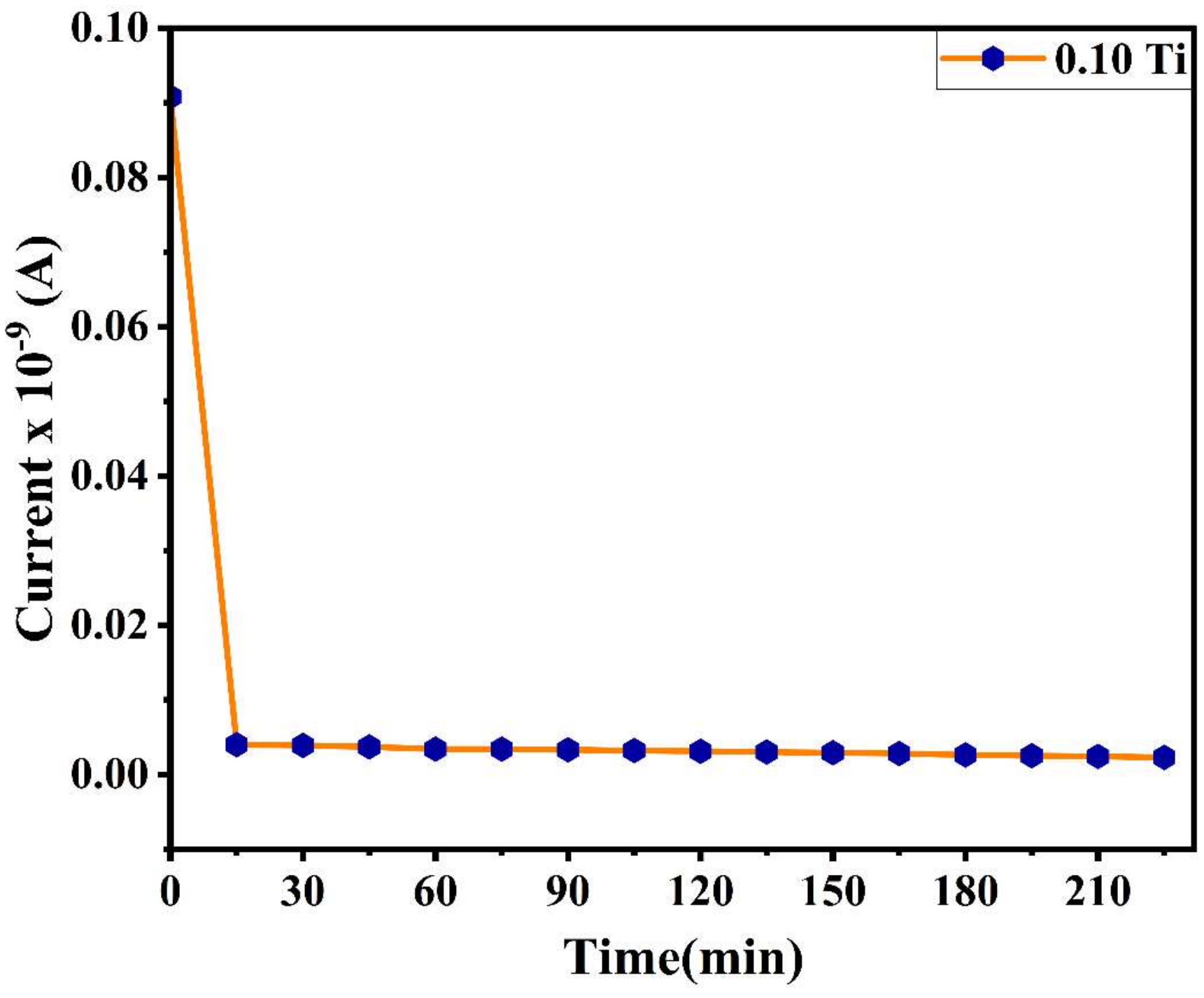

**Fig.8: DC conductivity graph for 0.10 Ti sample.**

## Conclusions:

The garnet type $Li_7La_3Zr_{2-x}Ti_xO_{12}$ (x= 0 - 0.20) series was prepared using the conventional solid state reaction method. The XRD analysis confirmed the formation of a conducting cubic phase within all the samples. Among all the samples, the 0.10 Ti possessed a dense microstructure with the maximum relative density of 91.8%. The obtained highest density value can be attributed to the sintering behaviour of Ti. The Elemental mapping revealed the homogeneous distribution of all the elements in the 0.10 Ti sample. The maximum ionic conductivity of $8.08\times10^{-5}Scm^{-1}$ was offered by the 0.10 Ti with a lowest activation energy of 0.37eV. This result can be attributed to the enhancement in Li-ion migration pathways owing to the compact microstructure of the sample. Moreover, the DC conductivity measurement verified the predominance of ionic conduction in the 0.10 Ti sample. Overall, as compared to pure LLZO, there is a significant increase in ionic conductivity by one order of magnitude. Hence, from the obtained results, it can be concluded that the 0.10 Ti sample can be a suitable candidate for further investigation in order to use it in the ASSLIB application.

**Acknowledgement:** Authors gratefully acknowledge the Department of Physics, VNIT Nagpur, for granting access to the X-ray diffraction (XRD) facility established under the DST-FIST project, grant number SR/FST/PSI/2017/5(C).

**Declaration of Interest Statement:** Authors declare no competing interest'

**Funding:** No external financial support was received for this research work from any funding agency in the public, commercial, or not-for-profit sectors.